\documentclass[10 pt, conference, twoside]{IEEEtran}

\usepackage{threeparttable}
\usepackage{amsmath,amssymb}
\usepackage{enumerate}

\newcommand{\beq}{\begin{equation}}
\newcommand{\beql}[1]{\begin{equation}\label{#1}}
\newcommand{\eeq}{\end{equation}}

\usepackage{etoolbox}
\IEEEoverridecommandlockouts
\usepackage{cite}
\usepackage{amsmath,amssymb,amsfonts}

\usepackage{tkz-euclide,tikzscale}
\usetikzlibrary{arrows,circuits.ee.IEC}
\tikzset{ac source/.style={
  circuit symbol lines,
  circuit symbol size = width 2 height 2,
  shape = generic circle IEC,
  /pgf/generic circle IEC/before background={
    \pgfpathmoveto{\pgfpoint{-0.8pt}{0pt}}
    \pgfpathsine{\pgfpoint{0.4pt}{0.4pt}}
    \pgfpathcosine{\pgfpoint{0.4pt}{-0.4pt}}
    \pgfpathsine{\pgfpoint{0.4pt}{-0.4pt}}
    \pgfpathcosine{\pgfpoint{0.4pt}{0.4pt}}
    \pgfusepath{stroke}
  },
  transform shape
}}

\usepackage{graphicx}
\usepackage{textcomp}
\usepackage{xcolor}
\usepackage{graphicx}      % include this line if your document contains figures
\usepackage{bm}
\usepackage{epsfig}
\usepackage{epstopdf}
\usepackage{float}
\usepackage{mathrsfs}
\usepackage{xtab}
\usepackage{color}
\usepackage{ifthen}
\usepackage{setspace}
\usepackage{algorithm}
\usepackage{algorithmic}
\usepackage{dsfont}
\usepackage{booktabs}
\usepackage{verbatimbox}

\usepackage{ntheorem}
\theoremseparator{:}
\newtheorem{theorem}{Theorem}

\usepackage[font=sc,labelsep=period]{caption}

\title{A Scenario-oriented Approach for Energy-Reserve \\Joint Procurement and Pricing\vspace{-0.1cm}}

\author{\IEEEauthorblockN{Jiantao Shi, Ye Guo*}
\IEEEauthorblockA{Tsinghua-Berkeley Shenzhen Institute\\
Tsinghua University\\
Shenzhen, China\\}
\and
\IEEEauthorblockN{Lang Tong}
\IEEEauthorblockA{School of Electrical \& Computer Engineering\\
Cornell University\\
Ithaca, NY, USA}
\and
\IEEEauthorblockN{Wenchuan Wu, Hongbin Sun}
\IEEEauthorblockA{Dept. of Electrical Engineering\\
Tsinghua University\\
Beijing, China}
\thanks{This work was supported in part by the National Science Foundation of China under Grant 51977115. 

Corresponding author: Ye Guo, e-mail: guo-ye@sz.tsinghua.edu.cn}}

    \makeatletter
\patchcmd{\@maketitle}
  {\addvspace{0.5\baselineskip}\egroup}
  {\addvspace{-1.2\baselineskip}\egroup}
  {}
  {}
\makeatother

\begin{document}
	\maketitle
	\begin{abstract}
	% for or in the basecase/non-base scenarios?
     We propose a scenario-oriented approach for energy-reserve joint procurement and pricing in electricity markets. In this model, without empirical reserve requirements, reserve is procured according to all possible contingencies and load/renewable generation fluctuations in non-base scenarios, and the deliverability of reserve is ensured through network constraints in all scenarios considered. Based on the proposed model, a locational marginal pricing approach has been developed for both energy and reserve. The associated settlement process is also discussed in detail. Under certain assumptions, the proposed pricing approach is a set of uniform pricing at the same location and the property of revenue adequacy for the system operator has also been established. 
	\end{abstract}

	\IEEEpeerreviewmaketitle

	\section{Introduction}
	{W}{ith} increasing concerns about energy shortages and global warming, integrating more renewable generations into the power system has become a worldwide trend. However, uncertain and intermittent renewable generations have brought new challenges to power systems' secure and reliable operations. To deal with these challenges, system operators need to properly procure reserve from dispatch-able resources to defend against possible contingencies and load/renewable generation fluctuations. Since energy and reserve are tightly coupled, all the independent system operators (ISOs) in the U.S. run joint optimizations \cite{review1}, by and large with the following model:
	\begin{align} 
	\label{obj1}
	& \mbox{(I)}: \quad\underset{\{g,r_U,r_D\}}{\rm minimize} \quad C_{g}^T g + C_{U}^T r_U + C_{D}^T r_D\\
	&\mbox{subject to}\notag\\
	\label{traditional balance}
	&\lambda : \mathds{1}^T g =  \mathds{1}^T d,\\
	\label{traditional pf}
	&\mu : S(g-d)\leq f,\\
	\label{reserve requirement}
	&(\gamma^U,\gamma^D):\mathds{1}^T r_U =  R^U,\mathds{1}^T r_D =  R^D,\\
	\label{physical limit}
	&g + r_U \leq \overline{G}, \underline{G} + r_D \leq g,0 \leq r_U \leq \overline{r_U}, 0 \leq r_D \leq \overline{r_D}.
	\end{align}
	The objective function (\ref{obj1}) aims to minimize the bid-in cost of energy and reserve with decision variables $(g,r_U,r_D)$, which denote vectors of energy, upward reserve, and downward reserve procured from generators. Coefficients $(C_{g},C_{U},C_{D})$ denote bid-in price vectors of $(g,r_U,r_D)$. Constraints (\ref{traditional balance})-(\ref{physical limit}) represent, respectively, energy balancing constraints, line capacity limits, reserve requirement constraints, and generator capacity and ramping rate limits. Reserve clearing prices will be obtained from the Lagrangian multipliers $(\gamma^U,\gamma^D)$, representing the increase of total bid-in cost when there is one additional unit of upward/downward reserve requirement. 
	
	Regarding the traditional co-optimization model (I), there are several key issues that remain unclear. The first is about reserve requirements $R^U$ and $R^D$. These parameters are artificial and empirical, sometimes specified as the capacity of the biggest generator as in PJM\cite{PJM} or a certain proportion of system loads as in CAISO\cite{CAISO}. However, they will significantly affect the clearing results and prices of reserve and even the energy market. The second is on the locations of reserve resources. The traditional model only considers network limits in the base case. When there are contingencies or load/renewable generation fluctuations, the deliverability of reserve may still be limited by line capacities. A common solution is to partition the entire system into different zones and specify zonal reserve requirements as in ISO-NE\cite{ISO-NE}, which is again empirical. The third is about the objective function (\ref{obj1}), which is to minimize the base-case bid-in cost but does not consider possible re-dispatch costs under contingencies and fluctuations.
	
	\indent In light of these problems, there are also many research efforts from academia. Some researches utilize the forecast methods, such as the density forecasts which consider the probability distributions of future observations\cite{MatosBessa1}, or the scenario forecasts which consider several typical future scenarios\cite{scen1}, to define reserve requirements in systems with high renewable penetrations, see \cite{requirementReview} for a survey. In \cite{reservezone}, a statistical clustering algorithm is proposed to enable dynamic reserve zone partition. In \cite{readjust1} and \cite{readjust2}, the energy balancing and network constraints considering events and re-dispatches in contingency scenarios are analyzed, and the reserve pricing approach at the bus level is proposed. In \cite{UMP}, the uncertainty marginal prices are defined to price both uncertainty sources and reserve resources at the bus level.
	
	\indent In this paper, a scenario-oriented energy-reserve co-optimization model is developed. These scenarios will correspond to possible contingencies or load/renewable generation fluctuations, or their combinations. Reserve procured from each generator is equal to the maximum range of its generation re-dispatch among all scenarios, thereby eliminating the need to specify the parameters of reserve requirements artificially. Transmission capacity limits in all scenarios are also included in the proposed model, therefore the artificial partition of reserve zones is no longer needed either. Thereupon, we derive marginal prices of generations, loads and reserve. Under certain assumptions, energy and reserve marginal prices will be locational uniform prices. We have also established the property of revenue adequacy for SOs: revenue from load payments, credits to generators including energy, reserve, and re-dispatches, and congestion rent will reach their balance in the base case as well as in all scenarios considered.

	\section{Energy-Reserve Co-optimization Model}
\begin{figure}[H]
		\centering
		\vspace{-0.35cm}
		\includegraphics[width=3.4in]{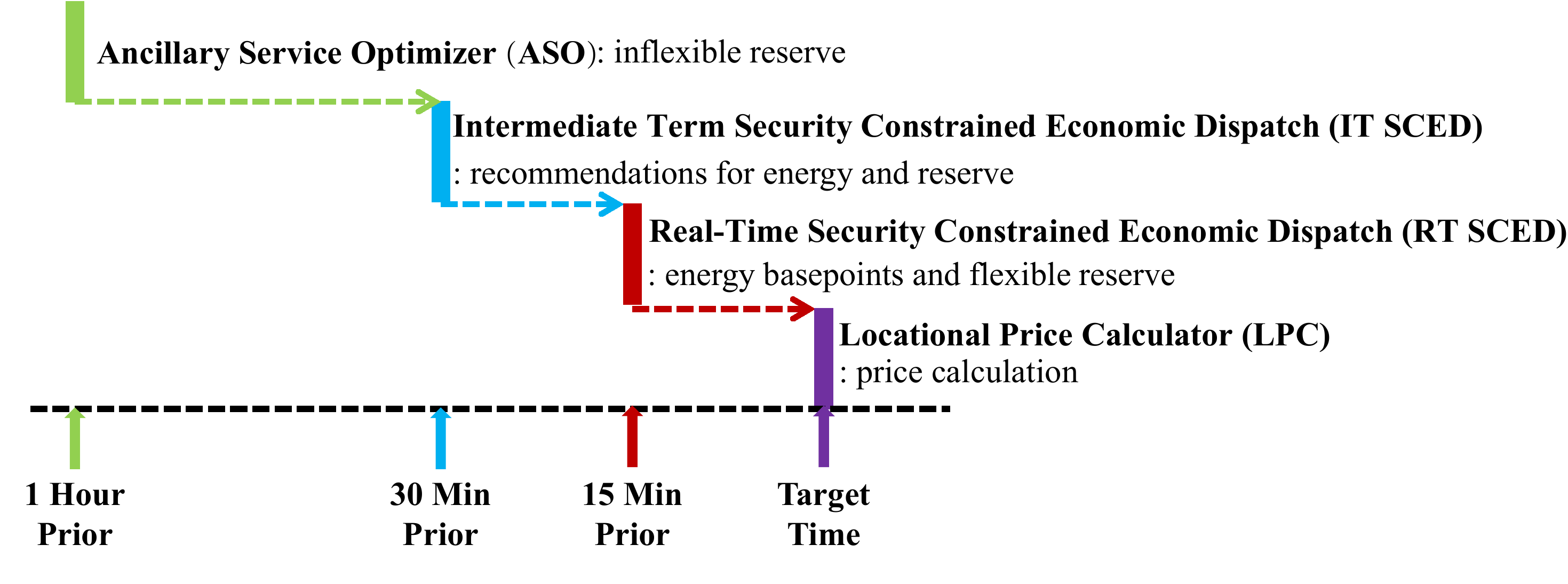}
    \setlength{\abovecaptionskip}{-3.0pt}
    \setlength{\belowcaptionskip}{-2.5pt}
    \captionsetup{font=footnotesize,singlelinecheck=false}
		\caption{PJM real-time operations including ASO, IT SCED, RT SCED and LPC\cite{PJM}}
		\label{PJM RT SCED}
    \vspace{-0.15cm}
	\end{figure}

     To properly address the problems of the traditional model (\uppercase\expandafter{\romannumeral1}), a scenario-oriented energy-reserve joint optimization model is proposed. Consider the current market structure, the proposed model can be regarded as an intermediate look-ahead optimization stage between the day-ahead (DA) scheduling and the real-time security constrained economic dispatch (RT SCED), like the ancillary service optimizer (ASO) in PJM. As Fig. \ref{PJM RT SCED} shows, ASO is one of many look-ahead stages to procure reserve resources with different flexibility levels. In practice, ASO is only financial binding for reserve\cite{PJM}. In this paper, however, we consider the proposed co-optimization model to be financial binding for both energy and reserve.
	
	The proposed model makes the following assumptions:
	
	1) A standard DCOPF model with linear cost functions for energy and reserve is adopted, which is consistent with the current electricity market design.
		
	2) A single-period problem is considered for simplicity\footnote{A multi-period setting will bring in the coupling between ramping and reserve, which is a highly complicated issue.}.
	
	3) For simplicity, we don't consider the DA market. If the DA market clearing results need to be considered, the quantities in the proposed model can be seen as the deviations from the quantities in the DA scheduling, and all the qualitative analyses in this paper will still hold.
		
	4) Uncertainties from line outages and load fluctuations are considered. Generator outages are not considered\footnote{Generator outages will bring in the complicated issue of non-uniform pricing. Essentially, energy and reserve procured from generators with different outage probabilities are no longer homogeneous goods. We will skip this tricky problem now and leave it to our future work.}. 
		
	5) Renewable generations are modeled as negative loads.
		
	6) The SO's predictions of possible scenarios as well as their occurrence probabilities are perfect.

	Based on assumptions (1)-(6), the proposed model is:
	\begin{equation}
	\setlength\abovedisplayskip{1pt}%shrink space
    \setlength\belowdisplayskip{-2.5pt}
    \begin{split}
	&\!F(g,r_U,r_D,\delta g^U_k,\delta g^D_k,\delta d_k)=\\
	\label{obj2}
	&\!C_{g}^T\!g\!+\!C_{U}^T\!r_U\!+\!C_{D}^T r_D\!+\!\sum^{k\in \mathcal{K}} \!\epsilon_k (\overline{C}^T_k\!\delta g^U_k \!-\!\underline{C}_k^T\!\delta g^D_k \!+\!C^T_{L}\!\delta d_k)\!,
    \end{split}
    \end{equation}
    \begin{align}
	&\mbox{(II)}: \quad \underset{\{g,r_U,r_D,\delta g^U_k,\delta g^D_k,\delta d_k\}}{\rm minimize} F(\cdot), \notag\\
	&\mbox{subject to}\notag\\	
	\label{base balance and pf}
	&(\lambda,\mu):\mathds{1}^T g =  \mathds{1}^T d, S(g-d)\leq f,\\
	\label{base physical limit}
	&g + r_U \leq \overline{G}, \underline{G} + r_D \leq g,0 \leq r_U \leq \overline{r_U}, 0 \leq r_D \leq \overline{r_D},\\
	&\mbox{for all $k\in \mathcal{K}$:} 
	\notag\\
	\label{cntg balance}
	&\lambda_k:\mathds{1}^T(g+\delta g^U_k-\delta g^D_k)=\mathds{1}^T (d+\pi_k -\delta d_k),\\
	\label{cntg pf}
	&\mu_k:S_k\bigl((g+\delta g^U_k-\delta g^D_k)-(d+\pi_k -\delta d_k)\bigr) \leq f_k,\\
	\label{dg1fanwei}
	&(\underline{\alpha_k},\overline{\alpha_k}): 0 \leq \delta g^U_k \leq r_U,\\
	\label{dg2fanwei}	
	&(\underline{\beta_k},\overline{\beta_k}): 0 \leq \delta g^D_k \leq r_D,\\
	\label{dk1fanwei}
	&(\underline{\tau_k},\overline{\tau_k}): 0 \leq \delta d_k \leq d+\pi_k,
	\end{align}
    where $\mathcal{K}$ denotes the set of all non-base scenarios, coefficient $\epsilon_k$ denotes the occurrence probability of scenario $k$, decision variables $(\delta g^U_k,\delta g^D_k,\delta d_k)$ denote vectors of generation upward re-dispatches, downward re-dispatches, and load shedding in scenario $k$. Coefficients $(\overline{C_k},\underline{C_k})$ represent vectors of generation upward and downward re-dispatch prices\footnote{In some ISOs in the U.S., generation re-dispatches will be settled at the real-time LMPs plus some premiums or adders e.g. ERCOT\cite{ERCOT}.} in scenario $k$. If one generator's downward reserve is deployed, it will pay back to the SO, so there is a negative sign before $(\underline{C}_k^T\delta g^D_k)$ in (\ref{obj2}). Coefficient $C_L$ denotes the vector of load shedding prices. Coefficient $\pi_k$ denotes the vector of load fluctuations in scenario $k$. The objective function (\ref{obj2}) aims to minimize the expected system total cost, including the base-case bid-in cost and the expectation of re-dispatch costs in all scenarios. Constraints (\ref{base balance and pf})-(\ref{base physical limit}) denote the base-case limits as constraints (\ref{traditional balance})-(\ref{traditional pf}) and (\ref{physical limit}) in the traditional model (I). Constraints (\ref{cntg balance})-(\ref{cntg pf}) denote the energy balancing constraints and the network constraints in all non-base scenarios where $S_k$ and $f_k$ denote the shift factor matrix and the line capacity vector in scenario $k$. In non-base scenarios, the power flows on transmission lines are allowed to exceed line capacities for short duration, and the exceeding ratings are reflected in the setting of $f_k$. Constraints (\ref{dg1fanwei})-(\ref{dg2fanwei}) denote that generation re-dispatches in all scenarios shall not surpass the procured reserve, indicating that the procured reserve will be modeled as the maximum range of generation re-dispatch among all scenarios. Therefore reserve will be optimally procured wherever necessary to minimize the expected total cost. Constraint (\ref{dk1fanwei}) denotes that load shedding in each scenario must be non-negative and cannot exceed actual load power in that scenario.

	With such formulations, the solution to the proposed model will procure reserve at the best locations considering all kinds of costs. On the contrary, the solution to the traditional model will be either sub-optimal or infeasible in order to minimize the expected total cost subject to energy balancing and line capacity constraints in both the base case and all scenarios, even with reserve requirements $R^U$ and $R^D$ being the sum of the optimal procurement $r_U^*$ and $r_D^*$ in model (II).
	
	Moreover, marginal prices derived from the proposed scenario-oriented optimization model (II) have many attractive properties, as in the next section.

	\section{Pricing and Settlement}
	Based on the proposed model, in this section we present the pricing approach for generations, loads, and reserve, as well as the associated market settlement process.
	\subsection{LMP calculations for generations, loads and reserve}
	The proposed pricing approach for energy and reserve is based on their marginal contributions to the expected system total cost presented in (\ref{obj2}). Namely, consider any generator $j$, we first fix $g(j),r_U(j),r_D(j)$ at their optimal values $g(j)^*,r_U(j)^*,r_D(j)^*$ and consider them as parameters instead of decision variables. Such a modified optimization model is referred to as model (III). Compared with the original model (II), in model (III) the terms that are only related to generator $j$, specifically the bid-in cost of generator $j$ in the objective function (\ref{obj2}) and the $j^{th}$ row of all constraints in (\ref{base physical limit}), will be removed. Next, we evaluate the sensitivity of the optimal objective function of model (III), which represents the expected cost of all other market participants except for generator $j$, with respect to the deviations in parameters $g(j),r_U(j)$ and $r_D(j)$. According to the envelop theorem, we have:
	\begin{equation}
	\setlength\abovedisplayskip{-0.5pt}%shrink space
    \setlength\belowdisplayskip{2pt}
	\frac{\partial C^*_{\uppercase\expandafter{\romannumeral3}}}{\partial g(j)}\!=\!\frac{\partial \mathcal{L}_{\uppercase\expandafter{\romannumeral3}}}{\partial g(j)}\!=\!S(:,m_j)^T\mu\!-\!\lambda\!+\!\sum^{k\in \mathcal{K}} (S_k (:,m_j)^T \mu_k\!-\!\lambda_k),
	\end{equation}
	where $C^*_{\uppercase\expandafter{\romannumeral3}}$ denotes the optimal objective function of model (\uppercase\expandafter{\romannumeral3}) and $\mathcal{L}_{\uppercase\expandafter{\romannumeral3}}$ denotes the Lagrangian function of model (\uppercase\expandafter{\romannumeral3}). Accordingly the marginal energy price of any generator is:
	\begin{align}
	\label{RGMP2}
	\eta^g(j)&\!=\!- \frac{\partial \mathcal{L}_{\uppercase\expandafter{\romannumeral3}}}{\partial g(j)}\!=\!\lambda\!-\!S(:,m_j)^T\!\mu\!+\!\sum^{k\in \mathcal{K}}\!(\lambda_k\!-\!S_k(:,m_j)^T\!\mu_k)\notag\\
	&=\omega_0(j)+\sum^{k\in \mathcal{K}}\omega_k(j),
	\end{align}
	where $\eta^g(j)$ denotes the marginal energy price of generator $j$, with the term $m_j$ denoting bus $m$ where generator $j$ is located. The term $(\lambda-S^T\mu)$, denoted by $\omega_0$ for brevity hereafter, denotes the base-case component of the energy prices of generators. The term $(\lambda_k-S_k^T\mu_k)$, denoted by $\omega_k$ for brevity hereafter, denote the non-base component of the energy prices of generators in scenario $k$. We can see that each component has an energy part and a congestion part, as in the standard LMP formulation. Similarly, the marginal energy prices of loads are calculated as
	\begin{align}
	\label{L/d}
	\eta^d&\!=\!\frac{\partial C^*_{\uppercase\expandafter{\romannumeral3}}}{\partial d}\!=\!\frac{\partial \mathcal{L}_{\uppercase\expandafter{\romannumeral3}}}{\partial d}\!=\!\lambda\!-\!S^T\mu\!+\!\sum^{k\in \mathcal{K}}(\lambda_k\!-\!S_k^T\mu_k)\!-\!\sum^{k\in \mathcal{K}}\overline{\tau}_k
	\notag\\&\!=\!\omega_0+\sum^{k\in \mathcal{K}}\omega_k-\sum^{k\in \mathcal{K}} \overline{\tau}_k,
	\end{align}
	which is consistent with the marginal energy prices of generators except for the last term $(-\sum \overline{\tau}_k)$, which are multipliers associated with the upper bound of load shedding in (\ref{dk1fanwei}).
	
	According to the envelop theorem, there is
	\begin{equation}
	\label{RUMP}
	\eta^U(j)=-\frac{\partial C^*_{\uppercase\expandafter{\romannumeral3}}}{\partial r_U(j)}=\frac{\partial\mathcal{L}_{\uppercase\expandafter{\romannumeral3}}}{\partial r_U(j)}=\sum^{k\in \mathcal{K}} \overline{\alpha_k}(j),
	\vspace{-0.25cm}
	\end{equation} 
	where $\eta^U(j)$ denotes the upward reserve marginal price of generator $j$. Similarly the downward marginal reserve price is:
	\begin{equation}
	\label{RDMP}
	\setlength\abovedisplayskip{-0.5pt}%shrink space
    \setlength\belowdisplayskip{1.5pt}
	\eta^{D}(j)=\sum^{k\in \mathcal{K}}\overline{\beta_k}(j).
	\end{equation}
	With such formulations, the connections between reserve and generation re-dispatches can be established: if the upward generation re-dispatch $\delta g^U_k(j)\!=\!r_U(j)$ in scenario k, then its corresponding multiplier $\overline{\alpha_k}(j)$ will be positive and contribute to the upward reserve marginal price $\eta^U(j)$.
	
	\indent Although the energy prices of generators and loads are defined separately, and the reserve marginal prices in (\ref{RUMP})-(\ref{RDMP}) are defined at the resource level, next we establish the uniform pricing property with some additional assumptions:
		
    7) Assume that $(-\sum\!\overline{\tau}_k)$ is zero in (\ref{L/d})\footnote{It is rare to shed a load to its total capacity. If a load is completely shed and another load at the same bus is not, this indicates that they have different reliability requirements, which will again bring in the non-uniform pricing issue.}.
	
	8) Assume that generators at the same bus will have the same $(\overline{C}_k,\underline{C}_k)$ in each scenario\footnote{The upward and downward re-dispatch prices $(\overline{C}_k,\underline{C}_k)$ in scenario $k$ can be set as the forecast RT-LMPs in scenario $k$ plus a fixed adder. With perfect predictions, the forecast RT-LMPs will be equal to the actual RT-LMPs, therefore the upward/downward re-dispatch prices $(\overline{C}_k,\underline{C}_k)$ can be used to settle the generation upward/downward redispatches.}.

	\begin{theorem}[Uniform Pricing]
	    Consider any two generators $i,j$ and any load $l$ at the same bus. Under assumptions (1)-(7), there is $\eta^g(i)=\eta^g(j)=\eta^d(l)$ for energy. 
	   
	   	Moreover, under assumptions (1)-(8) and assume that $r_U(i),r_D(i),r_U(j),r_D(j)\!>\!0$, there are $\eta^U\!(i)\!=\!\eta^U\!(j)$ for upward reserve and $\eta^D\!(i)\!=\!\eta^D\!(j)$ for downward reserve.
	\end{theorem}
	
	 The proof is presented in the Appendix A. Due to the page limit, we leave the proofs for this Theorem and the following Theorem 2 to the online version of our manuscript \cite{shi2020scenariooriented}. 
	 
	 The settlement process based on the proposed pricing approach has some attractive properties, as in the next subsection.
	
	\subsection{Market settlement process}
	Next the settlement process will be presented. The process can be separated into two stages. In the ex-ante stage, we don't know which prediction is true; and in the ex-post stage, the settlement depends on which scenario is actually realized.
	\subsubsection{ex-ante stage}
	 In this stage, generations and reserve will be financial binding and settled, also the basic load capacities $d$ and load fluctuations $\pi_k$ in all scenarios will be settled. Therefore, the ex-ante stage includes the following payments: 
	\begin{itemize}
		\itemsep=2.2pt
		\item contribution of base-case prices to generator energy credit: 
		\begin{equation}
		\label{base-case prices to generator energy credit}
		\setlength\abovedisplayskip{1pt}
        \setlength\belowdisplayskip{-0.5pt}
		\Gamma^g_0=\omega_0^Tg;
		\end{equation}
		
		\item contributions of non-base scenario prices to generator energy credit:	
		\begin{equation}
		\label{non-base scenario prices to generator energy credit}
		\setlength\abovedisplayskip{1pt}
        \setlength\belowdisplayskip{-0.5pt}
		\sum\Gamma^g_k=\sum^{k\in \mathcal{K}}\omega_k^Tg;
		\end{equation}
		
		\item contributions of base-case price to load energy payment:
		\begin{equation}
		\label{base-case prices to load payment}
		\setlength\abovedisplayskip{-0.5pt}
        \setlength\belowdisplayskip{-0.5pt}
		\Gamma^d_0 = (\omega_0)^Td;
		\end{equation}
		
		\item contributions of non-base scenario prices to load energy payment:
		\begin{equation}
		\label{non-base scenario prices to load payment in all scenarios}
		\setlength\abovedisplayskip{-2pt}
		\setlength\abovedisplayskip{-1pt}
		\sum^{k\in \mathcal{K}}\Gamma^d_k = \sum^{k\in \mathcal{K}}(\omega_k)^T d;
		\end{equation}
		
	\item non-base load fluctuation payment in all scenarios:
		\begin{equation}
		\label{fluctuation payment}
		\setlength\abovedisplayskip{1.6pt}
		\setlength\abovedisplayskip{-1pt}
			\sum^{k\in \mathcal{K}}\Pi_k = \sum^{k\in \mathcal{K}}(\omega_k)^T \pi_k;
		\end{equation}
		
	\item upward and downward reserve credit:
		\begin{equation}
		\label{upward reserve payments}
	    \setlength\abovedisplayskip{0.5pt}
		\Gamma^U=(\eta^U)^T r_U=\sum^{k\in \mathcal{K}}\overline{\alpha_k}^T r_U=\sum^{k\in \mathcal{K}}\Gamma^U_k,
		\end{equation}
		\begin{equation}
		\label{downward reserve payments}
		\setlength\belowdisplayskip{-0pt}
		\Gamma^D=(\eta^D)^T r_D=\sum^{k\in \mathcal{K}}\overline{\beta_k}^T r_D=\sum^{k\in \mathcal{K}}\Gamma^D_k.
		\end{equation}
	\end{itemize}
	
	For the fluctuation payment, note that one load should pay for its fluctuations in all scenarios because the SO will procure reserve to deal with all possible load fluctuations accordingly, therefore the load fluctuation payments should not only depend on the realized scenario in real-time, but should also rely on other scenarios considered.
    \subsubsection{ex-post stage}	 
	In this stage, one of the non-base scenarios is actually realized and denoted as scenario $k$. With perfect predictions, for each generator $j$, the deviation of its real-time generation level from its base-case energy procurement $g(j)^*$ will be $\delta g^U_k(j)$ or $\delta g^D_k(j)$. Under the assumption (7) and the corresponding footnote (5), upward and downward generation re-dispatches will be settled with $\overline{C_k}$ and $\underline{C_k}$, and load shedding will be settled with the shedding prices $C_L$. Therefore, the ex-post stage includes the following payments: 
	\begin{itemize}
	    \itemsep=1.0pt
		\item upward redispatch compensation:
		\begin{equation}
		\label{upward redispatch payments}
		\setlength\abovedisplayskip{-0.5pt}
		\setlength\belowdisplayskip{-1pt}
		\Phi^U_k\!=\!\overline{C}^T_k\delta g^U_k;
		\end{equation}
		\item downward redispatch pay-back:
		\begin{equation}
		\label{downward redispatch payments}
		\setlength\abovedisplayskip{-0.5pt}
		\setlength\belowdisplayskip{-1pt}
		\Phi^D_k\!=\!\underline{C}^T_k\delta g^D_k;
		\end{equation}
		\item load shedding credit:
		\begin{equation}
		\label{load shedding credits}
		\setlength\abovedisplayskip{-0.5pt}
		\setlength\belowdisplayskip{-1pt}
		\Phi^d_k\!=\!C_L ^T \delta d_k.
		\end{equation}
	\end{itemize}
	Note that the net revenue of the SO should be equal to the congestion rent. For the settlement process, the following theorem regarding SO's revenue adequacy is established: 
	\begin{theorem}[Revenue Adequacy]   
    Under assumptions (1)-(8), the net revenue of the SO is always non-negative.

	In particular, for the base case, load energy payment (\ref{base-case prices to load payment}) is equal to the sum of generator energy credit (\ref{base-case prices to generator energy credit}) and congestion rent:
	\begin{equation}
	\setlength\abovedisplayskip{-0.5pt}
	\setlength\belowdisplayskip{-0.5pt}
	\Gamma^d_0=\Gamma^g_0+\Delta_0=\Gamma^g_0+f^T\mu.
	\end{equation}
	
	For each non-base scenario k, its contribution to load payment is equal to the sum of its contribution to generator credit, load shedding credit, and congestion rent:
	\begin{equation}
	\setlength\abovedisplayskip{0.5pt}
	\setlength\belowdisplayskip{-0.5pt}
	\Gamma^d_k +\Pi_k=\Gamma^g_k+\Gamma^U_k+\Gamma^D_k+\epsilon_k\Phi^U_k -\epsilon_k\Phi^D_k+\epsilon_k\Phi^d_k+\Delta_k,k \in \mathcal{K},
	\end{equation}
	where the left-hand side is the contribution of scenario k to load payment, including energy payment (\ref{non-base scenario prices to load payment in all scenarios}) and load fluctuation payment (\ref{fluctuation payment}). The first five terms on the right-hand side represent its contribution to generator credit, including energy credit (\ref{non-base scenario prices to generator energy credit}), upward and downward reserve credit (\ref{upward reserve payments}-\ref{downward reserve payments}), upward and downward re-dispatch payment (\ref{upward redispatch payments}-\ref{downward redispatch payments}). The sixth term is the expected load shedding credit (\ref{load shedding credits}), and the last term is the congestion rent in that scenario $\Delta_k=f_k^T \mu_k$.
	\end{theorem}

    Please refer to the Appendix B in \cite{shi2020scenariooriented} for the proof. With this Theorem, payments from loads, payments to generators and congestion rent will reach their balance in the basecase as well as in all scenarios. Also, the reserve costs and the re-dispatch costs will be allocated in a scenario-oriented way.
	
	\section{Case Study}
    The case studies are performed both on a 2-bus system and on the modified IEEE 118-bus system.
	\subsection{Two-bus System}
	 \begin{figure}[H]
		\centering
		\captionsetup{font=footnotesize,singlelinecheck=false}
	    \vspace{-0.4cm}
		\includegraphics[width=2in]{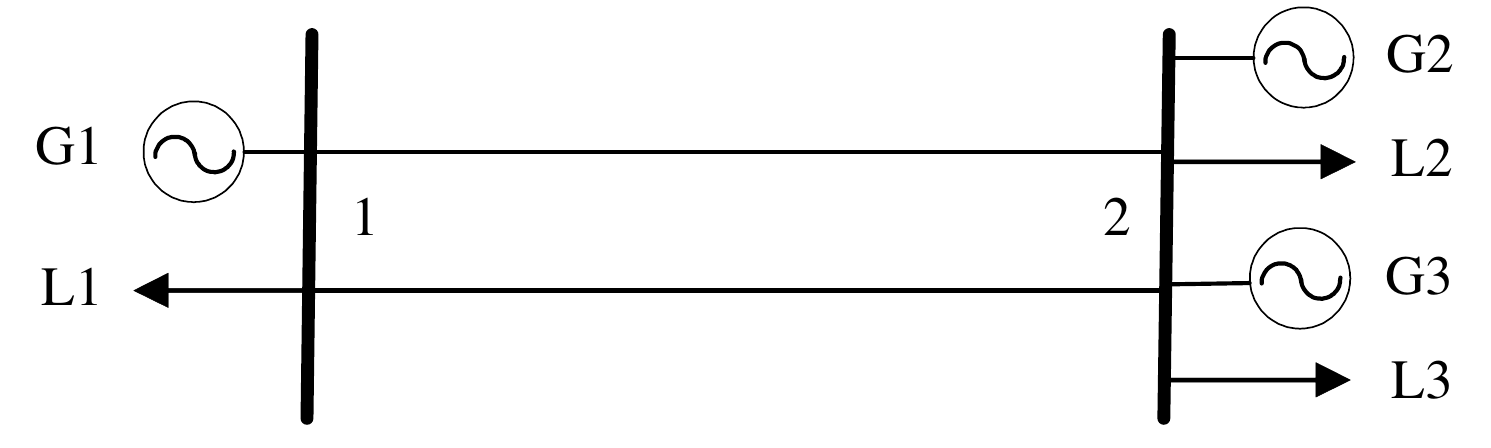}
	    %\vspace{-0.2cm}	
		\caption{the One-Line Diagram for the 2-Bus System}
		\label{one line diag}
	    \vspace{-0.3cm}	
	\end{figure}
    We first consider a 2-bus system with its one-line diagram presented in Fig. \ref{one line diag}. The generator bids are presented in Table \ref{Generators' Offers}. The outage probability of the two-parallel-line branch is 10\%, meaning that the system will lose one of these two lines if the outage happens. The line capacity exceeding rates are set to be for all scenarios. The basic loads are $(6,15,4)$MW, with two possible fluctuation situations: situation I $(+2,+6,-1)$MW with probability 20\%, and situation II $(+3,+2,-3)$MW with probability 20\%. Based on the line outage and load fluctuation information, we present all possible scenarios in Table \ref{Non-base}.
    
    	\begin{table}[H]
	    \footnotesize
		\centering
		\vspace{-0.2cm}
		\setlength{\abovecaptionskip}{-0.5pt}
        \setlength{\belowcaptionskip}{0pt}
        \captionsetup{font=footnotesize,textfont=sc}
		\caption{Generators' Offer Data for the 2-Bus System}
	%\vspace{-0.2cm}
		\begin{tabular}{cccc}
			\hline
			Generator &$\overline{G}/\underline{G}$& $\overline{r_U}/\overline{r_D}$& $C_{g}/C_{U}/C_{D}$ \\
			\hline
		 G1  & $16/0$ & $4/4$ & $8/2/2$\\
		 G2  &$18/0$  &  $4/4$ & $15/2/2$  \\
		 G3 & $12/0$  &  $4/4$ &  $20/2.5/2.5$   \\
			\hline
			%\bottomrule[1pt]
			\end{tabular}
		\label{Generators' Offers}
		\vspace{-0.2cm}
		\end{table}
		
	\begin{table}[H]
	\centering
	\vspace{-0.3cm}
	\footnotesize
    \setlength{\abovecaptionskip}{-0.5pt}
    \setlength{\belowcaptionskip}{0pt}
        \captionsetup{font=footnotesize,textfont=sc}
			\caption{Non-Base Scenario Data for the 2-Bus System}
			\label{Non-base}
			\begin{tabular}{ccccc}
				\hline
				NO.  & Outage & Load Situation & Probability & $\overline{C}$ and $\underline{C}$\\
				\hline
				1 & No & basic load & $0.06$ & $[19.1;26.3]$\\
				2 & Yes & situation I  & $0.02$& $[19.7;33.8]$  \\
				3 & Yes & situation II  & $0.02$& $[19.4;32.7]$  \\
				4 & No & situation I& $0.18$& $[19.4;33.5]$ \\
				5 & No & situation II& $0.18$& $[19.1;27.5]$ \\
				\hline
			\end{tabular}
		\vspace{-0.25cm}
		\end{table}
		
	\indent We present the clearing results in Table \ref{clear}. While G1 offers the cheapest upward reserve bid and still owns extra capacity/ramping rate, the SO doesn't clear its entire upward reserve bid, instead the more expensive resources G2 and G3 are cleared. The reason is that the extra upward reserve from G1 can't be delivered in scenarios when the branch outage happens. It can also be observed that there are uniform energy and reserve prices at each bus. In the meantime, the load fluctuation payments can calculated according to equation (\ref{fluctuation payment}): $\Pi^d(1)$=\$23.3, $\Pi^d(2)$=\$91.7, $\Pi^d(3)$=\$-23.5. Note that d3 fluctuation payment is negative because d3's fluctuations will hedge the fluctuations of d1 and d2.

		\begin{table}[H]
		\footnotesize
			\centering
        \vspace{-0.1cm}
        \setlength{\abovecaptionskip}{-0pt}
        \setlength{\belowcaptionskip}{0pt}
            \captionsetup{font=footnotesize,textfont=sc}
			\caption{Clearing Results for the 2-Bus System}	
			\label{clear}
			\begin{tabular}{ccccccc}
				%\toprule[1pt]
				\hline
				Generator & $g$ & $r_U$ & $r_D$& $\eta^g$ & $\eta^U$ & $\eta^D$\\
				\hline
				G1 & $8.0$ & $2.4$ & $0.8$& $25.4$ & $2.0$ & $2.0$ \\
				G2 & $17.0$ & $1.0$ & $0.0$& $35.7$ & $5.3$ & $3.7$ \\
				G3 & $0.0$ & $4.0$ & $0.0$& $35.7$ & $5.3$ & $3.7$ \\
				\hline
			\end{tabular}
		\vspace{-0.2cm}
		\end{table}
		
		\begin{table}
		\footnotesize
            \captionsetup{font=footnotesize,textfont=sc}         
            \caption{Money Flow for the 2-Bus System(\$)}
			\label{money flow}
			\centering
			\vspace{-0.2cm}
			\begin{tabular}{ccccccccc}
				\hline
				& Base & S1 & S2 & S3 & S4 & S5& Total\\
				\hline
				$\Gamma^d$ & $446.9$ & $24.9$ & $25.2$ &$9.7$ & $238.2$  &  $86.0$ & $830.7$\\
				$\Pi^d$\ & $0$ &  $0$ & $8.0$ & $0.8$ & $75.9$   & $6.9$& $91.4$\\
				$\epsilon\Phi^d$ & $0$ & $0$ & $2.1$  & $0$ & $6.4$  & $0$& $8.6$\\
				\hline
				$\Gamma^g$ & $443.4$& $20.0$ &  $23.6$  & $9.7$ & $227.5$  & $86.0$& $810.1$\\	
				$\Gamma^U$ & $0$ & $0$ & $2.6$ & $0$  & $28.7$  & $0$& $31.3$\\	
				$\Gamma^D$& $0$ & $1.6$ & $0$ & $0$ & $0$ & $0$ & $1.6$\\
				$\epsilon\Phi^U$ & $0$ & $1.3$ &$4.0$ &$0.9$ & $38.5$ & $7.5$ & $52.1$\\
				$\epsilon\Phi^D$ & $0$ & $0.9$ & $0.1$ & $0.1$ & $0$ & $0.6$ & $1.7$\\
				$\Delta$ & $3.5$& $2.9$ &$1.0$ &$0$& $12.7$ & $0$ & $20.1$\\
				\hline
			\end{tabular}
			\begin{tablenotes}
			\item[a] base / S1-S5: denote the base case / scenario 1-5;
			\item[b] Revenue adequacy: $(\Gamma^d=\Gamma^g+\Delta)$ holds for column 1, $(\Gamma^d\!+\!\Pi^d\!-\!\epsilon\Phi^d\!=\!\Gamma^g\!+\!\Gamma^U\!+\!\Gamma^D\!+\!\epsilon\Phi^U\!-\!\epsilon\Phi^D\!+\!\Delta)$ holds for other columns.
			\end{tablenotes}	
			\vspace{-0.3cm}
		\end{table}

	\indent In Table \ref{money flow}, the money flow is presented. Explicitly revenue adequacy holds in the base case, in each scenario and in total.

	\subsection{IEEE 118-Bus System}
	\indent Simulations on the modified IEEE 118-bus system are also reported. Outage probabilities of lines 21, 55, and 102 are set to be 10\%. The line capacity exceeding rates are set to be 1.2 for all scenarios. The original load 59 will be equally separated into two loads: new load 59 and load 119. There will be two load fluctuation situations, each with occurrence probability 10\%, and the fluctuation levels of all loads will be 3\% in both situations: in situation I, d119 will increase by 3\% while other loads will decrease by 3\%; in situation II, d119 will decrease by 3\% while others will rise by 3\%. The generators' energy and reserve bids are modified to be linear. The total revenue inadequacy in this case is $\$2.26$, which is rather small compared to the expected total cost $\$89476.4$.
	\begin{figure}[!t]
		%\vspace{-0.1cm}	
		\centering
		\includegraphics[width=3.9in]{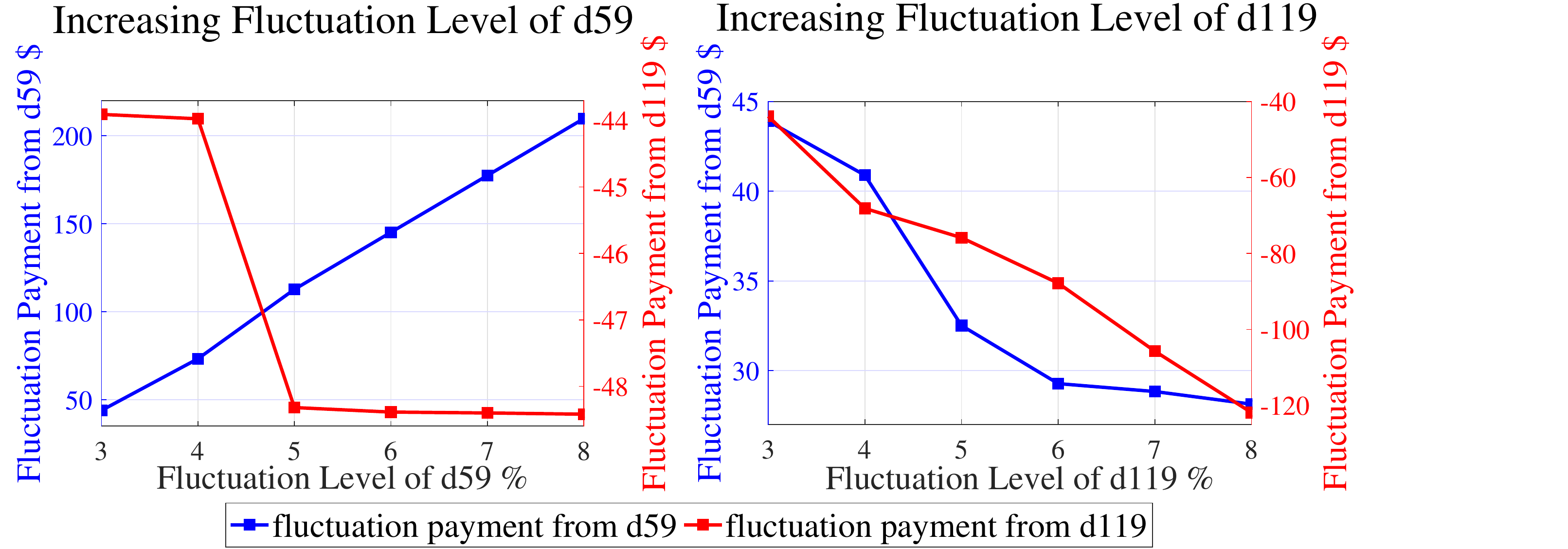}
		\vspace{-0.25cm}
        \setlength{\abovecaptionskip}{-1.5pt}
        \setlength{\belowcaptionskip}{-3.5pt}
		%坐标轴 16pt
		%label 17.6pt
		%legend 17.6pt
		%title 24pt
	    \captionsetup{font=footnotesize,singlelinecheck=false,}
		\caption{Fluctuation Payments from d59 (blue) and d119 (red) with the Fluctuation Levels of d59 (left) and d119 (right)}
		\label{plot_M_on_fluc_d_1_2}
		\vspace{-0.35cm}
	\end{figure}

	    %\begin{figure}[t]
		%\centering
		%\includegraphics[width=4in]{plot_Cplus_pi_on_u_b.eps}
		%\vspace{-0.2cm}
		%\includegraphics[width=3.9in]{plot_M_on_u_b.eps}
		%\vspace{-0.15cm}
		%坐标轴 16pt
		%label 17.6pt
		%legend 17.6pt
		%title 24pt
		%\vspace{-0.9cm}
		%\captionsetup{font=footnotesize,singlelinecheck=false}
		%\caption{Price of $r_U(7)$ with the Increasing Fluctuation Levels of d15 (left) and Price of $r_D(28)$ with the Increasing Fluctuation Levels of d66 (right)}
		%\label{plot_M_on_u_b}
		%\vspace{-0.6cm}
	%\end{figure}
	
     \indent In Fig. \ref{plot_M_on_fluc_d_1_2}, the fluctuation payment from d119 is negative because its fluctuation will hedge the others' fluctuations in all scenarios, so d119's fluctuations will be credited. For the left figure, With the rising fluctuation level of d59, the fluctuation payment from d59 will increase while the fluctuation credit to d119 will increase because d59's rising fluctuation level will bring more uncertainties, therefore enhancing the value of d119's fluctuations hedge against others' fluctuations. For the right figure, while the fluctuation credit to d119 will increase with its rising fluctuation level, fluctuation payment from d59 will decrease because the rising fluctuation level of d119 will reduce the impact of d59's fluctuations on system balance.

    %Fig. \ref{plot_M_on_u_b} depicts that with the increasing fluctuation level of d15, the upward reserve price of G7 at bus 15 will increase (left), and with the increasing fluctuation level of d66, the downward reserve price of G28 at bus 66 will increase (right). Underlining in these phenomenons is that the increasing fluctuation levels of d15 and d66 will bring in more uncertainties, therefore enhancing the value of reserve at the same bus.
    
    %We also compare the expected system total cost from the proposed model and the traditional model. For different base-case procurement levels $(g,r_U,r_D)$, we generate much future scenarios and calculate optimal re-dispatch costs in all scenarios under different base-case procurement levels, and consider the average of the re-dispatch costs as their corresponding expected re-dispatch costs. If the re-dispatch problems are infeasible, the re-dispatch costs for them will be the biggest re-dispatch cost among all feasible re-dispatch problems. 
	
		    \begin{figure}[t]
		\centering
		%\includegraphics[width=4in]{plot_Cplus_pi_on_u_b.eps}
		%\vspace{-0.2cm}
		\includegraphics[width=3.9in]{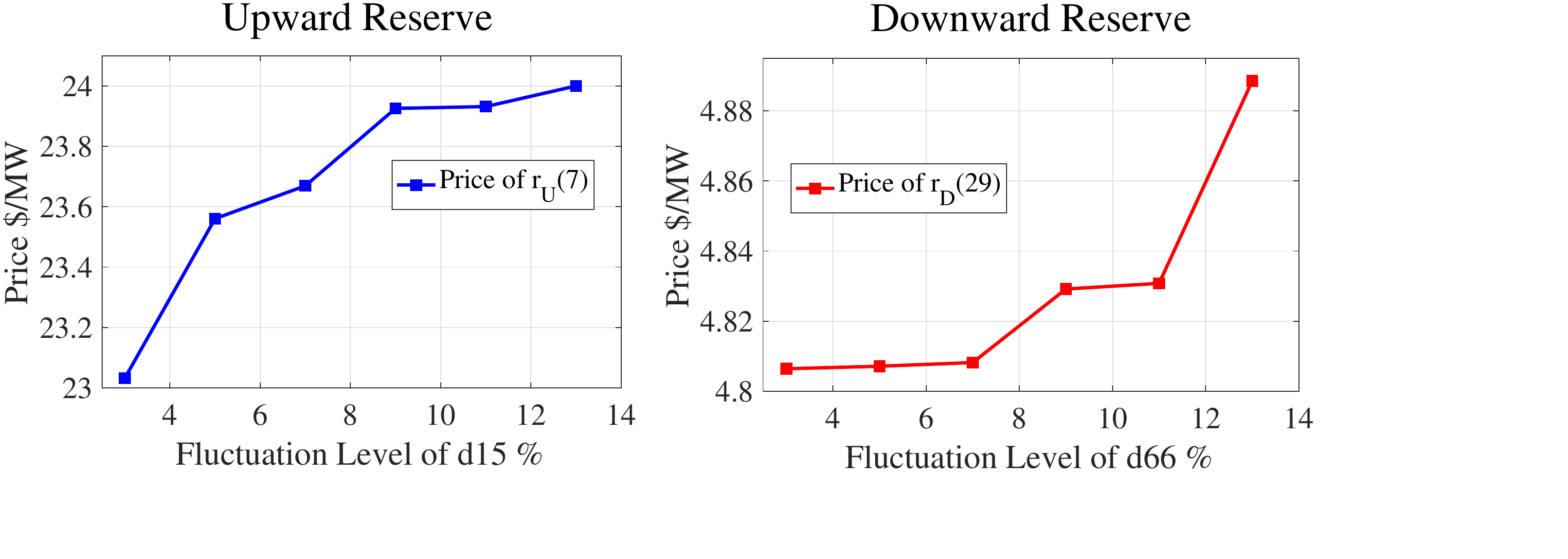}
		\vspace{-0.15cm}
		%坐标轴 16pt
		%label 17.6pt
		%legend 17.6pt
		%title 24pt
		\vspace{-0.9cm}
		\captionsetup{font=footnotesize,singlelinecheck=false}
		\caption{Price of $r_U(7)$ with the Increasing Fluctuation Levels of d15 (left) and Price of $r_D(28)$ with the Increasing Fluctuation Levels of d66 (right)}
		\label{plot_M_on_u_b}
		\vspace{-0.6cm}
	\end{figure}
	
    Fig. \ref{plot_M_on_u_b} depicts that with the increasing fluctuation level of d15, the upward reserve price of G7 at bus 15 will increase (left), and with the increasing fluctuation level of d66, the downward reserve price of G28 at bus 66 will increase (right). Underlining in these phenomenons is that the increasing fluctuation levels of d15 and d66 will bring in more uncertainties, therefore enhancing the value of reserve at the same bus.
    
	\section{Conclusions}
	In this paper, we proposed a scenario-oriented energy-reserve co-optimization model which considers the re-dispatch costs in all non-base scenarios to minimize the expected system total cost, and includes the network constraints in all scenarios to ensure the deliverability of reserve. We defined energy and reserve marginal prices which are locational uniform prices under certain assumptions, and proposed the associated settlement process which can guarantee revenue adequacy of the system operator. In future studies, we aim to include generator outages into the model, and consider the coupling between reserve and ramping in multi-period operations.
	
	\ifCLASSOPTIONcaptionsoff
	\newpage
	\fi

	
	
	%\bibliographystyle{IEEEtran}
	%\begin{thebibliography}{1}
	%\end{thebibliography}{1}
	\bibliographystyle{ieeetr}    %引用风格
	\bibliography{lib_test}	
	
		\appendices
	\renewcommand{\theequation}{\arabic{equation}}
	
	\section{Proof of Theorem 1}
	First consider the energy prices. For generators $i,j$ and load $l$ at the same bus $m$, their marginal prices will be:
	\begin{align}
	\label{uniform energy}
	&\eta^g(i)\!=\!\eta^g(j)\!=\!\lambda\!-\!S(:,m)^T\!\mu\!+\!\sum(\lambda_k\!-\!S_k(:,m)^T\!\mu_k),\\
	&\eta^d(l)=\lambda-S(:,m)^T\mu+\sum\lambda_k-\sum S_k(:,m)^T\mu_k,
	\end{align}
	apparently $(\eta^g(i)=\eta^g(j)=\eta^d(l))$ holds.
	
   To consider the reserve prices, we need to fundamentally analyze the relationships between the generation redispatches and procured reserve, so we can substitute the generation redispatches $\delta g^U_k,\delta g^D_k$ as follows:
	\begin{equation}
    \setlength\abovedisplayskip{0.5pt}
    \setlength\belowdisplayskip{-0pt}
	\delta g^U_k=x_k*r_U,\delta g^D_k=y_k*r_D,
	\end{equation}
	where $x_k$ and $y_k$ are both diagonal matrices. For each indicators $(x_k(j),y_k(j) \in [0,1], j=1,..N_g)$ on their diagonals, they represent the ratio of generator $j$'s upward/downward generation redispatch to generator $j$'s procured upward/downward reserve in scenario $k$. With these substitutions, we can transform the original model (\uppercase\expandafter{\romannumeral2}) into a new model (IV). In the new model, the optimal energy and reserve procurement and redispatches in all non-base scenarios won's change, also the corresponding Lagrangian multipliers in these two models will be the same. The new model is:
	 \begin{align}
	&F'(g,r_U,r_D,x_k,y_k,\delta d_k)=  C^T_{g} g + C^T_{U} r_U + C^T_{D} r_D\notag \\&+\sum\limits^K \epsilon_k (\overline{C}^T_k(x_k r_U)-\underline{C}_k^T(y_k r_D) +C^T_L\delta d_k),\notag\\
	 &(\uppercase\expandafter{\romannumeral4})\underset{\{g,r_U,r_D,x_k,y_k,\delta d_k\}}{\rm minimize} F'(\cdot), 
	\notag\\ &\mbox{subject to}
	\notag\\
	&(\ref{base balance and pf}),(\ref{base physical limit}),(\ref{dk1fanwei}),\mbox{for all $k \in \mathcal{K}$:}\notag\\
	\label{cntg balance1}
	&(\lambda_k)\mathds{1}^T(g+x_k r_U - y_k r_D)=\mathds{1}^T (d+\pi_k -\delta d_k),\\
	\label{cntg pf1}
	&(\mu_k)S_k\bigl((g + x_k r_U - y_k r_D)-(d+\pi_k -\delta d_k)\bigr) \leq f_k,\\
	\label{dg1fanwei1}
	&(\underline{\alpha_k},\overline{\alpha_k},\underline{\beta_k},\overline{\beta_k}) 0 \leq x_k r_U \leq r_U,0 \leq y_k r_D \leq r_D,
	\end{align}
	We denote the Lagrangian function of model (\uppercase\expandafter{\romannumeral4}) as $\mathcal{L}_{\uppercase\expandafter{\romannumeral4}}$. According to the KKT condition and ignoring $-  \underline{\tau_k}$, we have: 
	\begin{align}
	\label{L/a}
	&\frac{\partial \mathcal{L}_{\uppercase\expandafter{\romannumeral4}}}{\partial \delta d_k(l)}=
	 \epsilon_kC_{L}(l) 
	 -\lambda_k  +  S_k(:,m_l)^T\mu_k= 0,
	\end{align}
	\begin{align}
	\label{L/b}
	&\frac{\partial \mathcal{L}_{\uppercase\expandafter{\romannumeral4}}}{\partial x_k(j)} = 
    \epsilon_k\overline{C}_k(j) r_U(j) -\underline{\alpha_k}(j) r_U(j)  
	\notag\\& + \overline{\alpha_k}(j) r_U(j)- \lambda_k r_U(j) + S_k(:,m_j)^T\mu_k r_U(j)= 0,
	\end{align}
	\begin{align}
	\label{L/e}
	&\frac{\partial \mathcal{L}_{\uppercase\expandafter{\romannumeral4}}}{\partial y_k(j)} = -\epsilon_k\underline{C}_k(j) r_D(j)- \underline{\beta_k}(j) r_D(j)   
	\notag\\&+ \overline{\beta_k}(j) r_D(j) + \lambda_k r_D(j) - S_k(:,m_j)^T\mu_k r_D(j)= 0.
	\end{align}
   The assumption $(r^U(i),r^U(j)>0)$ indicates that $(\underline{\alpha_k}(i),\underline{\alpha_k}(j)=0)$, then according to equation (\ref{L/b}) we have:
    \begin{align}
    &\overline{\alpha_k}(i)=-\epsilon_k\overline{C}_k(i)+\lambda_k-S_k(:,m)^T\mu_k,\notag\\
    &\overline{\alpha_k}(j)=-\epsilon_k\overline{C}_k(j)+\lambda_k-S_k(:,m)^T\mu_k.\notag
	\end{align}
	Since we assume $\overline{C}_k(i)=\overline{C}_k(j)$, then apparently $(\overline{\alpha_k}(i)=\overline{\alpha_k}(j))$ holds, so we have:
    \begin{equation}
    \label{uniform upward}
    \setlength\abovedisplayskip{0.5pt}
    \setlength\belowdisplayskip{-0pt}
	\sum^K\overline{\alpha_k}(i)=\sum^K\overline{\alpha_k}(j),
    \end{equation}
	therefore generator $i$ and generator $j$ will receive the same upward reserve marginal prices.
	
	In the meantime, the assumption $(r^D(i),r^D(j)>0)$ indicates that $(\underline{\beta_k}(i),\underline{\beta_k}(j)=0)$ holds, then according to equation (\ref{L/e}) we have:
	\begin{align}
	\overline{\beta_k}(i) =\epsilon_k\underline{C}_k(i)-\lambda_k +S_k(:,m)^T\mu_k,\notag\\
    \overline{\beta_k}(j) =\epsilon_k\underline{C}_k(j)-\lambda_k +S_k(:,m)^T\mu_k.\notag
	\end{align}
	Since we assume $\underline{C}_k(i)=\underline{C}_k(j)$, then apparently $\overline{\alpha_k}(i)=\overline{\alpha_k}(j)$ holds, so we have:
    \begin{align}
    \label{uniform downward}
    \setlength\abovedisplayskip{0.5pt}
    \setlength\belowdisplayskip{-0pt}
	\sum^K\overline{\beta_k}(i)=\sum^K\overline{\beta_k}(j),
    \end{align}
	therefore generator $i$ and generator $j$ will receive the same downward reserve marginal prices. With equation (\ref{uniform energy}), (\ref{uniform upward}) and (\ref{uniform downward}) we can prove Theorem 1.

	\section{Proof of Theorem 2}
	If we multiply $\delta d_k(l),x_k(j),y_k(j)$ to both the left-hand side and the right-hand side of equations (\ref{L/a})-(\ref{L/e}), respectively, then with the complementary slackness of (\ref{dg1fanwei1}) we have:
	\begin{equation}
	\setlength\abovedisplayskip{-0.5pt}
    \setlength\belowdisplayskip{-0.5pt}
	\label{L/a1}
	 \lambda_k \delta d_k(l) =\epsilon_kC_{L}(l) \delta d_k(l)
	  + S_k(:,m_l)^T\mu_k \delta d_k(l),
	\end{equation}
	\begin{align}
	\label{L/b1}
	\overline{\alpha_k}(j)r_U(j)&=- \epsilon_k\overline{C}_k(j)x_k(j)r_U(j) + \lambda_k x_k(j)r_U(j) 
	\notag\\& \quad -S_k(:,m_j)^T\mu_k x_k(j) r_U(j),
	\end{align}
	\begin{align}
	\label{L/e1}
	\overline{\beta_k}(j)r_D(j)&= \epsilon_k\underline{C}_k(j)y_k(j)r_D(j) -  \lambda_k y_k(j)r_D(j)
	\notag\\& \quad + S_k(:,m_j)^T\mu_k y_k(j) r_D(j).
	\end{align}
	
	In the meantime, to consider the congestion rent, the phase angled based form of model (\uppercase\expandafter{\romannumeral4}) will be adopted as follows:
	\begin{align}
	&(\uppercase\expandafter{\romannumeral5}) \underset{\{g,r_U,r_D,x_k,y_k,\delta d_k,\theta,\theta_k\}}{\rm minimize} F'(\cdot), 
	\notag\\ &\mbox{subject to}\notag\\
	&(\ref{traditional balance}),(\ref{traditional pf}),(\ref{physical limit}),(\ref{dk1fanwei}),(\ref{dg1fanwei1})\\
	\label{basecase balance2}
	&(\Lambda) g-d = B \theta,\\
	\label{basecase pf2}
	&(\mu) F \theta \leq f,\\
	&\mbox{for all $k \in \mathcal{K}$:} \notag\\
	\label{cntg balance2}
	&(\Lambda_k)(g+x_k r_U - y_k r_D)- (d+\pi_k -\delta d_k)=B_k \theta_k,\\
	\label{cntg pf2}
	&(\mu_k)F_k \theta_k \leq f_k,
	%(\ref{basecase balance1}) (\ref{basecase pf1}) (\ref{cntg balance1}) (\ref{cntg pf1})
	\end{align}
	With the equivalence of the shift factor based model and the phase angle based model, we have:
    \begin{equation}
    \setlength\abovedisplayskip{-0.5pt}%shrink space
    \setlength\belowdisplayskip{-0.5pt}
	\label{cntgptdf}
	\Lambda= \lambda - S^T\mu,\Lambda_k = \lambda_k - S_k^T\mu_k, k \in \mathcal{K}.
    \end{equation}
	In the meantime, We denote the Lagrangian function of model (\uppercase\expandafter{\romannumeral5}) as $\mathcal{L}_{\uppercase\expandafter{\romannumeral5}}$, with the KKT condition we have:
	\begin{align}
	\label{baseCR}
	&\theta^T\frac{\partial \mathcal{L}_{\uppercase\expandafter{\romannumeral5}}}{\partial  \theta}=(B\theta)^T\Lambda + (F\theta)^T\mu = 0,\\
	\label{cntgCR}
	&\theta_k^T\frac{\partial \mathcal{L}_{\uppercase\expandafter{\romannumeral5}}}{\partial  \theta_k}=(B_k\theta_k)^T\Lambda_k + (F_k\theta_k)^T\mu_k = 0.
	\end{align}
	In the basecase, we have $(\Gamma^d_0-\Gamma^g_0=(\lambda - S^T\mu)^T(d-g))$ and $(\Delta_0=f^T\mu)$. To consider revenue adequacy in the basecase, we have:
	\begin{align}
	\label{lemma4}
	(\lambda\! - \!S^T\mu)^T(d\!-\!g)&\!=\!\Lambda^T(d\!-\!g)\!=\!-\!\Lambda^T(B\theta)\!=\!(F\theta)^T\mu\!=\!f^T\mu.
	\end{align}
    These four equations are based on the equation (\ref{cntgptdf}), the KKT conditions for constraint (\ref{basecase balance2}), the equation (\ref{baseCR}), and the KKT conditions for constraint (\ref{basecase pf2}), respectively. These equations indicate that $(\Gamma^d_0=\Gamma^g_0+\Delta_0)$, which proves revenue adequacy of the SO in the basecase.
	
	\indent In the meantime, the congestion rent contributed from any non-base scenario $k$ will be:
	\begin{align}
	\label{cntgCR2}
	&\quad f_k^T\mu_k = (F_k\theta_k)^T\mu_k =-(B_k\theta_k)^T\Lambda_k
	\notag\\&=\Lambda_k^T((d+\pi_k d -\delta d_k)-(g+x_k r_U-y_k r_D))
	\notag\\&=(\lambda_k - S_k^T\mu_k)^T((d+\pi_k d -\delta d_k)-(g+x_k r_U-y_k r_D))
	\notag\\&=(- S_k^T\mu_k)^T(d+\pi_k-\delta d_k)\!-\!(- S_k^T\mu_k)^T(g\!+\!x_k r_U\!+\!y_k r_D)
	\notag\\&=(- S_k^T\mu_k)^T(d+\pi_k)+(S_k^T\mu_k)^T\delta d_k +(S_k^T\mu_k)^T g
	\notag\\&\quad+(S_k^T\mu_k)^T(x_k r_U)-(S_k^T\mu_k)^T(y_k r_D).
	\end{align}
	These six equations are based on the KKT conditions for constraint (\ref{cntg pf2}), the equation (\ref{cntgCR}), the KKT conditions for constraint (\ref{cntg balance2}), the equation (\ref{cntgptdf}), the KKT conditions for constraint (\ref{cntg balance1}), and the reorganization of the equation, respectively. Also with the KKT conditions of constraint (\ref{cntg balance1}) and the equation (\ref{L/a1}) we have:
	\begin{align}
	&\quad\lambda_k \sum_j(g(j)+x_k(j) r_U(j)-y_k(j)r_D(j)))
	\notag\\&=\sum\limits_l \lambda_k (d(l)+\pi_k(l) -\delta d_k(l))
	\notag\\&\!=\!\sum\limits_l(\lambda_k d(l)\!+\!\lambda_k \pi_k(l) \!-\!\epsilon_k C_{L}(j) \delta d_k(l)\!-\!S_k(:,m_l)^T\epsilon_k \delta d_k(l)),\notag
	\end{align}
	which can be reorganized according to (\ref{L/b1})-(\ref{L/e1}) as follows:
	\begin{align}
	\label{RA2}
	&\quad	\sum\limits_l(\lambda_k d(l)+\lambda_k \pi_k(l) -\epsilon_kC_L(l)\delta d_k(l)
	\notag\\&\quad-S_k(:,m_l)^T \mu_k \delta d_k(l))-\sum\limits_j\lambda_k g(j)
	\notag\\&=\sum\limits_j (\overline{\alpha}_k(j)+\epsilon_k \overline{C}_k(i) x_k(j)+S_k(:,m_j)^T \mu_k x_k(j))r_U(j)
	\notag\\&\quad-\!\sum\limits_j (\!-\overline{\beta}_k(j)\!+\!\epsilon_k \underline{C}_k(j)y_k(j)\!+\!S_k(:,m_j)^T \mu_k y_k(j))r_D(j).
	\end{align}
    If we add $(\sum\limits_l\!S_k(:,m_l)^T\! \mu_k(d(l)\!+\!\pi_k(l)) \!+\! \sum\limits_j\!S_k(:,m_j)^T\!\mu_kg(i))$ and its opposite to the right hand side of (\ref{RA2}) and reorganize the equation, we have:
	\begin{align}
	\label{RA3}
	&\quad\sum\limits_l(\lambda_k-S_k(:,m_l)^T\mu_k))(d(l)+\pi_k(l))
	\notag\\&=\sum\limits_j(\lambda_k g(j)-S_k(:,m_j)^T \mu_k g(j))	
	\notag\\&\quad+\sum\limits_j\overline{\alpha}_k(j)r_U(j)+\sum\limits_j \overline{\beta}_k(j)r_D(j)
	\notag\\&\quad\!+\!\sum\limits_j \epsilon_k \overline{C}_k(j)x_k(j)r_U(j)\!-\!\sum\limits_j \epsilon_k \underline{C}_k(j)y_k(j)r_D(j)
	\notag\\&\quad\!-\!\sum\limits_l \epsilon_k C_L(l) \delta d_k(l)
	\notag\\&\quad +(\sum\limits_j S_k(:,m_j)^T \mu_k g(j) + \sum\limits_j S_k(:,m_j)^T \mu_k x_k(j)r_U(j)
	\notag\\&\quad - \sum\limits_j S_k(:,m_j)^T \mu_k y_k(j)r_D(j) + \sum\limits_l S_k(:,m_l)^T \mu_k \delta d_k(l)) \notag\\&\quad - \sum\limits_l S_k(:,m_l)^T\mu_k(d(l)+\pi_k(l)).
	\end{align}
    The term on the left-hand side of equation (\ref{RA3}) is the contribution of scenario k to load payment, including energy payment (\ref{non-base scenario prices to load payment in all scenarios}) and load fluctuation payment (\ref{fluctuation payment}). The right-hand side of equation (\ref{RA3}) include the energy credit (\ref{non-base scenario prices to generator energy credit}) in the $1^{st}$ row, upward and downward reserve credit (\ref{upward reserve payments}-\ref{downward reserve payments}) in the $2^{nd}$ row, expected upward and downward re-dispatch payment (\ref{upward redispatch payments}-\ref{downward redispatch payments}) in the $3^{rd}$ row, expected load shedding credit (\ref{load shedding credits}) in the $4^{th}$ row, the congestion rent in the $5^{th}$-$7^{th}$ rows which can be reorganized as equation (\ref{cntgCR2}). Therefore equation (\ref{RA3}) can also be written as:
	\begin{equation}
	\setlength\abovedisplayskip{0.5pt}
	\setlength\belowdisplayskip{-0.5pt}
	\Gamma^d_k +\Pi_k=\Gamma^g_k+\Gamma^U_k+\Gamma^D_k+\epsilon_k\Phi^U_k -\epsilon_k\Phi^D_k+\epsilon_k\Phi^d_k+\Delta_k,
	\end{equation}
    which can prove revenue adequacy of the SO in each scenario $k$. With equation (\ref{lemma4}) and (\ref{RA3}), we can prove Theorem 2. %bib文件名，同时该语句确定了参考文献出现的位置.
	
	
	%\vfill
	

\end{document}